\documentclass[leqno,12pt]{article}

\usepackage{amsmath}
\usepackage{amssymb}
\usepackage[authoryear]{natbib}
\usepackage{graphicx}
\usepackage{epsfig}
\usepackage{color}
\usepackage{setspace}

\makeatletter\@addtoreset{equation}{section}\makeatother

\newcommand{\R}{\mathbb{R}}

\newcommand{\argmin}{\operatorname*{argmin}}

\newcommand{\bea}{\begin{eqnarray*}}
\newcommand{\eea}{\end{eqnarray*}}
\newcommand{\be}{\begin{eqnarray}}
\newcommand{\ee}{\end{eqnarray}}
\newcommand{\beq}{\begin{equation}}
\newcommand{\eeq}{\end{equation}}
\newtheorem{assumption}{Assumption}[section]

\newtheorem{thm}{Theorem}[section]

\newtheorem{lemma}[thm]{Lemma}

\newtheorem{exam}[thm]{Example}
\newtheorem{algorithm}[thm]{Algorithm}

\def\3{\ss}

\DeclareMathOperator*{\arginf}{arg\,inf}
\DeclareMathOperator*{\argmax}{arg\,max}

\begin{document}

\title{Efficient computation of Bayesian optimal  \\ discriminating designs}

\author{
{\small Holger Dette} \\[-6pt]
{\small Ruhr-Universit\"at Bochum} \\[-6pt]
{\small Fakult\"at f\"ur Mathematik} \\[-6pt]
{\small 44780 Bochum, Germany} \\[-6pt]
{\small e-mail: holger.dette@rub.de}\\
{\small Roman Guchenko, Viatcheslav B. Melas } \\[-6pt]
\small St. Petersburg State University \\[-6pt]
\small Department of Mathematics \\[-6pt]
\small St. Petersburg,  Russia \\[-6pt]
{\small e-mail: vbmelas@post.ru,romanguchenko@ya.ru}}
\date{}
\maketitle

\begin{abstract}
An efficient algorithm for the determination of Bayesian optimal discriminating designs for competing regression models is developed, where the main focus
is on models with general distributional assumptions beyond the ``classical'' case of  normally distributed homoscedastic errors.
For this purpose we consider a Bayesian version of the Kullback-Leibler (KL) optimality criterion introduced by  \cite{loptomtra2007}.
Discretizing  the prior distribution leads to local KL-optimal discriminating design  problems for
 a large number of competing models.
All currently available methods either require a large computation time or 
fail to calculate the optimal discriminating design, because they can only deal 
efficiently with a few model comparisons. In this paper we develop a new algorithm  for the determination of
Bayesian optimal discriminating designs with respect to the Kullback-Leibler criterion. It is demonstrated that the new algorithm is able to calculate the 
 optimal discriminating designs  with reasonable accuracy and computational time in situations where all currently available procedures 
are either slow or fail.
 \end{abstract}


Keyword and Phrases: Design of experiment;  Bayesian optimal design; model discrimination; gradient methods; model uncertainty; Kullback-Leibler distance


\section{Introduction} \label{sec1}

Although optimal designs can provide a substantial improvement in the statistical accuracy without making any additional experiments,
classical optimal design theory [see  for example   \cite{pukelsheim2006,atkinson2007}] has been criticized, because it relies heavily
on the specification of a particular model. In many cases a good design for a given model might be inefficient if it is used in a different
setup.  Most of the literature addressing   the problem of model uncertainty in the design of experiments can be roughly divided into two parts, where
all authors assume that a certain
class of parametric models is available to describe the relation between the predictor and the response.
One approach to  obtain model robustness  is to construct   designs which allow the efficient estimation of parameters
in all models under consideration. This is usually achieved  by optimizing composite optimality criteria, which are defined as an average of the criteria for the different models
 [see \cite{laeuter1974}, \cite{dette1990,biedetpep2006,debrpepi2008}]. Alternatively,
 one can directly   construct designs to  discriminate between several competing models.   An early reference is \cite{stigler1971} who
determined  designs for discriminating between
two nested univariate polynomials by  minimizing  the volume of the confidence
ellipsoid for the parameters corresponding to the extension of the smaller model.
Since this seminal paper several authors have followed  this line of research
[see for example \cite{dethal1998} or  \cite{songwong1999} among others].
A completely different approach for the  construction of  optimal designs for model discrimination
was suggested  by \cite{atkfed1975a}. The corresponding optimality criterion is called  $T$-optimality criterion.
To be precise, assume that the relation between the response $Y$ and predictor $x$ is described by a
nonlinear regression  model such that
\be
\mathbb{E}[Y|x]  =\eta(x,\theta) ~, ~\mbox{Var}(Y|x) = v^2(x,\theta)~,   \label{1.1}
  \ee
and that the experimenter considers two rival models,
say $ \eta_1,  \eta_2$,  as candidates for the parametric form of the mean.
Roughly speaking,   \cite{atkfed1975a} assumed homoscedasticity, fixed one model, say $ \eta_1$, and    constructed the design
such that the sum of squares for a lack of fit test   against the alternative $\eta_2$ is large.
The criterion was extended in several directions. For example,
\cite{atkfed1975b} considered the problem of discriminating a selected model $\eta_1$ from
a class of other regression models,
say $ \{\eta_2, \ldots , \eta_\nu \}$, $\nu \ge 2$, and  \cite{Tommasi09}  combined the $T$-criterion with the approach introduced by
\cite{laeuter1974}.
 \cite{ucibog2005}   remarked that the criterion introduced by  \cite{atkfed1975a} is only applicable in the case of homoscedastic
 errors in the regression model \eqref{1.1} and discussed an  extension to  the case of
 heteroscedasticity. More generally,
  \cite{loptomtra2007}
introduced a generalization of the $T$-optimality criterion which is applicable under general distributional assumptions
and  called KL-optimality criterion. Meanwhile the determination of KL-optimal discriminating designs has
been discussed by several authors [see \cite{Tommasi09,tomlop2010}  among others]. \\
It is important to note here that the  $T$-optimality criterion and its extensions are   local optimality criteria in the sense of \cite{chernoff1953},
 because they require   the explicit knowledge of  the parameters in the model $\eta_1$.
As a consequence, optimal designs with respect to the $T$-optimality criterion might be sensitive with respect to misspecification of the parameters [see \cite {detmelshp2012} for a striking example].  A standard approach to obtain robust designs [which was already mentioned by \cite {atkfed1975a}] is the use of a Bayesian $T$-optimality criterion. This criterion is defined as an expectation of various local $T$-optimality criteria with respect to a prior distribution. 
\cite{detmelshp2012}  derived some explicit    Bayesian $T$-optimal designs   for
  polynomial regression models, but in general these designs have to be found numerically in nearly all cases of practical interest. 
  Recently,   \cite{detmelguc2015} pointed out that
  the numerical construction of Bayesian $T$-optimal designs is  an extremely difficult optimization problem, because
  -- roughly speaking -- the Bayesian optimality criterion corresponds to an optimal design problem for model discrimination
   with an extremely large number of competing models. As a consequence, the commonly used algorithms for the calculation of optimal designs,
   such as exchange-type methods or multiplicative methods and their extensions,
   cannot be applied to determine the Bayesian $T$-optimal discriminating design
   in reasonable computational time.  \cite{detmelguc2015} proposed a new algorithm
   for the calculation of Bayesian $T$-optimal discriminating designs and demonstrated its efficiency in several numerical examples.
   A drawback of this   method   consists still in the fact that it is only applicable to the ``classical''
   Bayesian $T$-optimality criterion which refers to the nonlinear regression model \eqref{1.1} with    homoscedastic and normally distributed responses, i.e. $\mathbb{P}^{Y|x} \sim \mathcal{N}(\eta(x,\theta), v^2(x,\theta))$.

The purpose of the present paper is to extend the methodology introduced by \cite {detmelguc2015} to regression models with more general
distributional assumptions. In Section \ref{sec2} we will introduce a Bayesian KL-optimality criterion which extends the criterion introduced by \cite{loptomtra2007}
to address for uncertainty in the model parameters. The criterion has also been discussed in \cite{tomlop2010}, who considered only two competing regression models. The new algorithm is
 proposed in Section \ref{sec3} and combines some features of the classical exchange type algorithms with gradient methods and quadratic programming. In Section \ref{sec4} we illustrate the applicability of the new method in several examples. In particular,
 we determine optimal discriminating designs with respect to the Bayesian KL-optimality criterion in situations where all other methods fail to find the optimal design. Finally, the appendix contains a proof of an auxiliary result.

\section{KL-optimal discriminating designs} \label{sec2}
The regression model \eqref{1.1} is a special case of a more general model, where the distribution of the random variable $Y$
has a density, say $f(y,x)$, and $x$ denotes an explanatory variable, which varies in a  compact
design space  $\mathcal{X}$. We assume that observations at different experimental conditions are
independent.   Following \cite{kiefer1974} we consider approximate
designs that are defined as probability measures, say $\xi$,  with finite support.
 The support points $ x_1,\ldots, x_k $ of a design $\xi$ give the locations
 where observations are taken, while the weights $\omega_1,\ldots, \omega_k$ describe the
 relative proportions of observations at these points. If an approximate design is given and  $n$
 observations can be taken, a rounding procedure is applied
to obtain  integers $n_{i} $ ($i=1,\ldots,k)$  from the not necessarily integer valued quantities
$\omega_{i}n$ such that $\sum_{i=1}^kn_i=n$.

Assume that the experimenter
wants to choose  a most appropriate model  from a given class, say
 $\{f_1, \ldots , f_\nu\}$ of competing models, where
   $f_j(y,x,\theta_j)$ denotes the density of the $j$th model with respect to a sigma-finite
 measure, say  $\mu$. The parameter $\theta_j$ varies in a compact parameter space
 $\Theta_j \ (j=1,\dots,\nu)$. The models may contain additional nuisance parameters,
 which will not be displayed in our notation. For two competing models, say $f_i$ and $f_j$, we denote by
 \begin{equation} \label{KL1}
 I_{i,j} (x,\theta_i, \theta_j) =  \int f_i (y,x,\theta_i) \log
 \frac {f_i (y,x, \theta_i)}{f_j(y,x,\theta_j)} \mu (dy)
 \end{equation}
 the  Kullback-Leibler distance between  $f_i$ and $f_j$.  If the model $f_i$
 is assumed to be the ``true'' model with parameter $\overline \theta_i$, then \cite{loptomtra2007}
 defined a local  KL-optimal discriminating design for the models $f_i$ and $f_j$ as a design maximizing the optimality criterion
 \begin{equation} \label{KL2}
\mathrm{KL}_{i,j} (\xi, \overline \theta_i) = \inf_{\theta_{i,j}\in \Theta_j}
\int_{\mathcal{X}} I_{i,j} (x, \overline \theta_i, \theta_{i,j}) \xi (dx).
 \end{equation}

  This  criterion can now easily be extended to construct optimal discriminating designs for more than two competing models.
  Following \cite{tomlop2010} and  \cite{BraessDette2013} we denote by $p_{i,j}$  nonnegative weights reflecting the importance
  of the comparison between the the model $f_i$ and $f_j$, where $f_i$ is assumed as the ``true'' model. The
(symmetrized) KL-optimality criterion  for more than $\nu \ge 2$ competing models $f_1,\dots,f_\nu$ is then defined by
\begin{align} \label{2.4}
\mathrm{KL}_{\mathrm{P}}(\xi)  = \sum_{i,j=1}^{\nu} p_{i,j} \mathrm{KL}_{i,j}(\xi , \overline{\theta}_i)
  =  \sum_{i,j=1}^{\nu} p_{i,j}   \inf_{\theta_{i,j} \in \Theta_j} \int_{\mathcal{X}}    I_{i,j} (x, \overline{\theta}_i,\theta_{i,j})
\xi(dx) ,
   \end{align}
   and a design maximizing the criterion \eqref{2.4} is called  local $\mathrm{KL}_{\mathrm{P}}$-optimal discriminating design for the models $f_1,\ldots, f_\nu$.
  For a design $\xi$ we also introduce the notation
\begin{align}
{\Theta}_{i,j}^*(\xi)  =   \arginf_{\theta_{i,j} \in \Theta_j} \int_{\mathcal{X}}
 I_{i,j} (x, \overline{\theta}_i,\theta_{i,j})   \xi(dx) ,
\label{thetamin}
\end{align}
 Our first result characterizes    local  KL-optimal discriminating design  and will be helpful to check the optimality of the numerically
 constructed designs. Its proof can be obtained by standard arguments and is therefore omitted.

\begin{thm} \label{thm1}
Let 
\begin{itemize}
\item[]
\begin{assumption}  \label{assum1}
For each   $i=1,\dots,\nu $ the function   $f_i(\cdot ,\cdot ,\theta_{i})$ is  continuously differentiable with respect to the  parameter $\theta_{i} \in \Theta_i,$
\end{assumption}
\end{itemize}

be satisfied.
A design  $\xi^*$ is a local  $\mathrm{KL}_\mathrm{P}$-optimal discriminating design, if and only if there
exist    distributions $\rho_{ij}^*$ 
 on the  sets $ {\Theta}_{i,j}^*(\xi^*)$ defined in  \eqref{thetamin} such that the
inequality
\begin{equation} \label{equiv}
 \sum_{i,j = 1}^{\nu} p_{i,j}  \int_{{\Theta}_{i,j}^*(\xi^*)}  I_{i,j} (x, \overline{\theta}_i,\theta_{i,j})   \rho_{ij}^* (d {\theta}_{i,j}) ~\leq 
  \mathrm{KL}_{\mathrm{P}}(\xi^*) 
 \end{equation}
 is satisfied for all
$ x \in \mathcal{X}$. Moreover, there is equality in \eqref{equiv}
 for all support points  of    the  local    $\mathrm{KL}_\mathrm{P}$-optimal discriminating design  $\xi^*$. \\
\end{thm}

If, additionally,
\begin{itemize}
\item[]
\begin{assumption}  \label{assum2}
 For any design $\xi$   such that
 $\mathrm{KL}_\mathrm{P}(\xi)>0$  and weight $p_{i,j} \neq 0$ the infima in \eqref{2.4} are attained at a unique points  $ \widehat{\theta}_{i,j}  = \widehat{\theta}_{i,j}(\xi)$
in the interior of the set $\Theta_j$,
\end{assumption}
\end{itemize}
is satisfied, then all measures $ \rho_{ij}^* $ in Theorem \ref{thm1}  are one-point measures and the left-hand side of inequality  \eqref{equiv}
simplifies to
\begin{equation} \label{psi}
\Psi(x,\xi) = \sum_{i,j = 1}^{\nu} p_{i,j}  I_{i,j} (x, \overline{\theta}_i,\widehat{\theta}_{i,j}) .
\end{equation}
Consequently, if $\xi$ is not a local $\mathrm{KL}_\mathrm{P}$-optimal discriminating design, it  follows that there exists
a point $\bar x \in \mathcal{X}$ such that $\Psi(\bar  x,\xi) > \mathrm{KL}_{\mathrm{P}}(\xi)$. \\

Note that the criterion \eqref{2.4} depends on the unknown parameters $\overline {\theta}_1, \ldots, \overline{\theta}_\nu$, which have to be specified by the experimenter for the competing model $f_1,\dots,f_\nu$, respectively.
Therefore the criterion is a local one in the sense of \cite{chernoff1953}. It was pointed out by \cite{detmelshp2012}   that the optimal designs maximizing the criterion \eqref{2.4} are rather sensitive with respect to misspecification of these parameters.
 For this reason we will now propose a Bayesian version of the criterion in order to obtain robust discriminating 
  designs for the competing models $f_1, \ldots , f_\nu$. \\
 We denote by $\mathcal{P}_i$ a prior distribution for the parameter $\overline \theta_i$ in model $f_i$ ($i=1,\ldots , \nu$) and define a Bayesian KL-optimality criterion by
 \begin{align}
\label{KLcriterion_general}
\mathrm{KL}_{\mathrm{P}}^{\mathrm{B}}(\xi) & = \sum_{i,j = 1}^{\nu} p_{i,j} \int_{\Theta_i}   \mathrm{KL}_{i,j} (\xi, \overline{\theta}_i) \mathcal{P}_i(d  \overline{\theta}_i), \\
& =\sum_{i,j = 1}^{\nu} p_{i,j} \int_{\Theta_i}  \inf_{\theta_{i,j} \in \Theta_j} \int_{\mathcal{X}}    I_{i,j} (x, \overline{\theta}_i,\theta_{i,j})
\xi(dx) \mathcal{P}_i(d \overline{\theta}_i) \nonumber
\end{align}
Optimal designs maximizing this criterion
 will be called Bayesian KL-optimal discriminating designs throughout this paper. We also note that the criterion  \eqref{KLcriterion_general}
  has been considered before by    \cite{tomlop2010} in the case of two competing regression models.  \\
It was pointed out by \cite{detmelguc2015} that the determination of Bayesian optimal discriminating designs with respect to the criterion
 \eqref{KLcriterion_general}
  is closely related to the problem of finding   local optimal discriminating designs for a large class of competing regression models.
  To be precise, we note that in most applications  the integral in \eqref{KLcriterion_general} is
 evaluated by numerical integration approximating  the prior distribution by a measure with finite support.
 Consequently,  if the prior distribution $ \mathcal{P}_{i}$ in the criterion is given  by a discrete measure with
masses $\tau_{i1}, \ldots \tau_{i\ell_i}$ at the points $\lambda_{i1},\ldots ,\lambda_{i\ell_i}$
the  criterion in \eqref{KLcriterion_general} can
   be  represented as
 \begin{align} \label{2.6}
\mathrm{KL}_{\mathrm{P}}^{\mathrm{B}}(\xi) = \sum_{i,j=1}^{\nu} \sum_{k=1}^{\ell_i} p_{i,j}  \tau_{ik} \inf_{\theta_{i,j} \in \Theta_j} \int_{\mathcal{X}}
  I_{i,j} (x, {\lambda_{ik}},\theta_{i,j})
 \xi(dx)  .
\end{align}
which is a local KL-optimality criterion of the from \eqref{2.4}, where the competing models are given by $\{ f_i (y,x, \lambda_{ik})|~  k=1,\dots,\ell_i; \ i=1,\dots,\nu \}$. The only difference between the criterion obtained from the (discrete) Bayesian approach
and  the criterion \eqref{2.4} consists in the fact that - due to discretization of the prior distributions $\mathcal{P}_1,\ldots  , \mathcal{P}_\nu$ -
 the criterion \eqref{2.6} involves substantially more comparisons of competing models $f_i(y,x,\lambda_{ik})$.
As a consequence the computation of Bayesian KL-optimal discriminating
design is computationally very challenging, because for each support point of the prior distribution in the criterion
 \eqref{2.6}  the infimum has to be calculated numerically.  In the following section we will propose  several new algorithms
 to address this problem. In Section \ref{sec4} it will be demonstrated that
 these methods yield very satisfactory results in cases where commonly used algorithms
 are either very slow or  fail to determine the Bayesian KL-optimal discriminating design.

\section{Efficient algorithms for  Bayesian KL-optimal designs }\label{sec3}

In this section we propose several algorithms for the calculation of Bayesian KL-optimal designs, which determine the optimal designs with reasonable accuracy and are computationally very
efficient. As pointed out in Section \ref{sec2} the Bayesian optimality criterion with
a discrete prior distribution reduces to
a local KL-optimality criterion of the form \eqref{2.4} with a large number of model comparisons. For this reason
we will describe the numerical procedures in this section for the criterion \eqref{2.4}. It is straightforward to extend the algorithms to the Bayesian criterion \eqref{2.6} and in  the following Section \ref{sec4} we
will give some illustrations  determining Bayesian KL-optimal discriminating designs by the new methods.

Most of the algorithms proposed in the literature for the calculation of optimal designs are based on the fact which was mentioned in the paragraph following Theorem \ref{thm1}. More precisely,
recall the definition of the function $\Psi$ in \eqref{psi} and assume that the design $\xi$
is not a Bayesian KL-optimal discriminating design. It then  follows under Assumption  \ref{assum2}
 that there exists  a point  $\overline{x} \in \mathcal{X}$, such that the inequality
$$
\Psi(\overline{x},\xi) > \mathrm{KL}_{\mathrm{P}}(\xi)
$$
holds.
 \cite{loptomtra2007} used this property to extend the algorithm of \cite{atkfed1975a} to the KL-optimality criterion.
 In the case of the  local KL-optimality criterion \eqref{2.4} it reads as follows.

\begin{algorithm}
\label{algorithm:AtkinsonFedorov}
Let $ \xi_0$ denote a given (starting) design and let $( \alpha_s)_{s=0}^{\infty}$   be  a sequence of positive numbers, such that
 $\lim_{s \to \infty} \, \alpha_s = 0, \; \sum_{s = 0}^{\infty} \alpha_s = \infty, \; \sum_{s = 0}^{\infty} \alpha_s^2 < \infty. $
For $s=0,1,\ldots $ define
$$\xi_{s+1} = ( 1 - \alpha_s ) \xi_s + \alpha_s \xi(x_{s+1}),$$
 where
$ x_{s+1} = \argmax_{x \in \mathcal{X}} \Psi(x,\xi_s).$
\end{algorithm}
\noindent
It can be shown that this algorithm yields a sequence of designs $(\xi_s)_{s \in \mathbb{N}}$ converging in the sense that
$\lim_{s \to \infty}  \mathrm{KL}_{\mathrm{P}}(\xi_s) =  \mathrm{KL}_{\mathrm{P}}(\xi^*)$, where $\xi^*$ denotes a local KL-optimal discriminating design.
However, it turns out that the rate of convergence is very slow. In particular, if there are many models under consideration, the algorithm 
is very slow and  fails 
in some models to determine the local KL-optimal discriminating design (see  our numerical example in Section \ref{sec4}). One reason for these difficulties consists in the fact that
Algorithm \ref{algorithm:AtkinsonFedorov}  usually yields a sequence of designs
with an  increasing number  of  support points. As a consequence  the resulting design (after applying some
stopping criterion)  is concentrated on a large set of points. In the case of normal distributed responses it is also demonstrated by  \cite{BraessDette2013}  that  Algorithm \ref{algorithm:AtkinsonFedorov} requires
a large number of  iterations if it is used for the  calculation of local KL-optimal discriminating designs
for more than two competing models.

Following \cite{detmelguc2015}  we therefore  propose an  alternative  procedure for the calculation of local  KL-optimal discriminating designs,
which  separates the maximization with respect to the support points  and weights in two steps.
In the discussion below we will present two methods for the calculation of the weights in
 the second step [see Section \ref{sec31} and \ref{sec32} for details].
\begin{algorithm}{\ }
\label{algorithm:new}{\rm
Let ${\xi_0}$ denote a starting design such that  $\mathrm{KL}_{\mathrm{P}} ({\xi_0})>0$ and define recursively
a sequence of designs $({\xi_s})_{s=0,1,\ldots } $  as follows:
\begin{itemize}
\item[$(1)$]  Let $\mathcal{S}_{[s]}$ denote the support of the design ${\xi_s}$. Determine
the set $\mathcal{E}_{[s]}$ of all local maxima of  the function $\Psi(x,{\xi_s})$ on the design space
 $\mathcal{X}$ and define
$\mathcal{S}_{[s+1]} =  \mathcal{S}_{[s]}  \cup \mathcal{E}_{[s]}$.
\item[$(2)$] Define $ \xi = \{\mathcal{S}_{[s+1]},\omega\} $ as the design supported at $ \mathcal{S}_{[s+1]}$ (with a normalized vector $w$ of non-negative weights) and determine the local  $\mathrm{KL}_{\mathrm{P}}$-optimal
design in the class of all  designs supported at $ \mathcal{S}_{[s+1]}$. In other words: we determine
the vector $\omega_{[s+1]}$ maximizing the function
\begin{align*}
g(\omega) =  \mathrm{KL}_{\mathrm{P}} ( \{\mathcal{S}_{[s+1]},\omega\}) = \sum_{i,j=1}^{\nu} p_{i,j} \inf_{\theta_{i,j} \in \Theta_j}
\sum_{x \in  \mathcal{S}_{[s+1]}}  I_{i,j} (x, \overline{\theta}_i,\theta_{i,j})
  w_x
\end{align*}
(here $w_x$ denotes the weights at the point $x \in \mathcal{S}_{s+1}$).
 All points in ${\mathcal{S}}_{[s+1]}$ with vanishing components in the vector of weights   $\omega_{[s+1]}$  will be
 be removed and the new set of support points will also be denoted by ${\mathcal{S}}_{[s+1]}$.
 Finally the design ${\xi}_{s+1}$ is defined as the design with the set of support points  ${\mathcal{S}}_{[s+1]}$ and the corresponding nonzero weights.
\end{itemize}
}
\end{algorithm}

\noindent
It follows by similar arguments as given in  \cite{detmelguc2015}  that the
sequence   $({\xi_s})_{s=0,1,\ldots } $  of designs generated by   Algorithm \ref{algorithm:new}
converges to  a   local KL-optimal discriminating  design. The crucial step in this algorithm is the second
one, because it requires -- in particular if a large number of competing models are under consideration --
the calculation of numerous infima. In order to address this problem we propose a quadratic programming and a
gradient method in the following two subsections.

\subsection{Quadratic programming} \label{sec31}
Let  $\mathcal{S}_{[s+1]} = \{x_1,\ldots , x_n\}$ denote the set
obtained in the first step of Algorithm \ref{algorithm:new} and recall the definition of the Kullback-Leibler distance $I_{i,j}$ in \eqref{KL1}. In Step 2 of Algorithm \ref{algorithm:new} a design  $\xi$  with
masses $\omega_1,\ldots , \omega_n$  at  the points $x_1,\ldots , x_n$ has to be determined such that
the function
\begin{align*}
g(\omega) = \sum_{i,j=1}^{\nu} p_{i,j}  \sum_{k=1}^n \omega_k \int  \log \Big\{\frac{f_i(y,x_k, \overline{\theta}_i)}{f_j(y,x_k, \widehat{\theta}_{i,j} )} \Big\} f_i(y,x_k,\overline{\theta}_i) d\mu(y).
\end{align*}
is maximal, where
\begin{align}\label{thetatilde}
 \widehat{\theta}_{i,j} = \widehat{\theta}_{i,j}  (\omega) =  \arginf_{\theta_{i,j} \in \Theta_j}  \sum_{k=1}^n \omega_k \int
   \log \left\{\frac{f_i(y,x_k,\overline{\theta}_i)}{f_j(y,x_k,{\theta}_{i,j})} \right\} f_i(y,x_k,\overline{\theta}_i) ~.
\end{align}
Define
\begin{align*}
\mathbf{J}_{i,j}(y) &= \mathbf{J}_{i,j}(\widehat{\theta}_{i,j},y) = \left( \frac{\sqrt{f_i(y,x_k,\overline{\theta}_i)}}{f_j(y,x_k,\widehat{\theta}_{i,j})}
\frac{\partial f_j(y,x_k,\theta_{i,j})}{\partial \theta_{i,j}}\Big|_{{\theta}_{i,j} = \widehat{\theta}_{i,j}}   \right)_{k=1}^n \in \mathbb{R}^{n \times d_j}, \\
\mathbf{R}_{i,j} &= \mathbf{R}_{i,j}(\widehat{\theta}_{i,j}) = \left( \int \frac{\partial f_j(y,x_k,\theta_{i,j})}{\partial
\theta_{i,j}}\Big|_{{\theta}_{i,j} = \widehat{\theta}_{i,j}}   \frac{f_i(y,x_k,\overline{\theta}_i)}{f_j(y,x_k,\widehat{\theta}_{i,j})} d\mu (y) \right)_{k=1}^n \in \mathbb{R}^{n \times d_j},
\end{align*}
and consider a linearized version of the function $g$, that is
\begin{align} \label{gbar}
\overline{g}(\omega) =
 \sum_{i,j=1}^{\nu} p_{i,j}
  \min_{\alpha_{i,j}} \sum_{k=1}^n \omega_k
   \left\{
    \int \log
     \left\{
      \frac{f_i(y,x_k,\overline{\theta}_i)}{f_j(y,x_k,{{\widehat{\theta}_{i,j}})}}
     \right\}
     f_i(y,x_k,\overline{\theta}_i) d \mu (y)
   \right. \\ \notag
+
   \left. \alpha_{i,j}^T (\mathbf{R}_{i,j}^T)_k - \frac{1}{2} \int \alpha_{i,j}^T (\mathbf{J}^T_{i,j}(y))_k (\mathbf{J}_{i,j}(y))_k  \alpha_{i,j} d \mu (y) \right\}.
\end{align}
Note that the minimum with respect to the parameters $\alpha_{i,j} \in \R^{d_j}$ is achieved for
\begin{align*}
 \widehat{\alpha}_{i,j} = \left( \int \mathbf{J}_{i,j}^T(y) \mathbf{\Omega} \mathbf{J}_{i,j}(y) d \mu (dy)
\right)^{-1} \mathbf{R}_{i,j}^T  \omega,
 \end{align*}
where the matrix $\mathbf{\Omega}$ is defined by $\mathbf{\Omega}=\mbox{diag}(\omega_1,\dots,\omega_n)$
 and $\omega= (\omega_1,\dots,\omega_n)^T$.
 For the following discussion we define  by $\Delta = \{ \omega \in \mathbb{R}^n ~|~\omega_i  \geq 0 ~(i=1,\ldots ,n) \; \sum_{i=1}^n \omega_i = 1
\}$ the simplex in $\R^n$.
\begin{lemma} \label{lem1}
If Assumptions \ref{assum1} and  \ref{assum2} are satisfied, then
each maximizer of the function $g(\cdot )$  in $\Delta$ is a maximizer of $\overline{g}(\cdot )$  in $\Delta$ and vice versa. Moreover,
\begin{align*}
\max_{\omega \in \Delta } g(\omega) = \max_{\omega \in \Delta} \overline{g}(\omega).
\end{align*}
\end{lemma}

\bigskip

A proof of Lemma \ref{lem1} can be found in the Section \ref{sec5}. With the notations
\begin{align*}
\mathbf{b}_{i,j} = \mathbf{b}_{i,j}({\widehat{\theta}_{i,j}}) = \left( \int \log \left\{\frac{f_i({y,x_k},\overline{\theta}_i)}{f_j({y,x_k},{\widehat{\theta}_{i,j}})} \right\} f_i({y,x_k},\overline{\theta}_i) 
{\mu(dy)} \right)_{k=1}^n { = \left( I_{i,j} (x_k,\overline{\theta}_i, \widehat{\theta}_{i,j}) \right)_{k=1}^n} ~\in \R^n
\end{align*}
we have
\begin{align*}
\overline{g}(\omega) = \mathbf{b}^T \omega -\omega^T \mathbf{Q}(\omega)\omega 
\end{align*}
where the vector $\mathbf{b} \in \mathbb{R}^n$ and the $n \times n$ matrix $\mathbf{Q}$ are  defined by
\begin{align*}
\mathbf{Q}(\omega) = \mathbf{R}_{i,j}\left(\int \mathbf{J}_{i,j}^T(y)\mathbf{\Omega}(\omega)\mathbf{J}_{i,j}(y){\mu(dy)}\right)^{-1} \mathbf{R}_{i,j}^T,
\end{align*}
 and $\mathbf{b} = \Sigma^\nu_{i,j=1} p_{i,j} \mathbf{b}_{i,j}$,
respectively. If we ignore the dependence of the matrix $\mathbf{Q}  (\omega ) $   and consider this matrix  as fixed
for a given matrix $\mathbf{\Omega} =
\mathrm{diag}(\overline{\omega}_1,\dots,\overline{\omega}_{n})$,
we obtain  a quadratic programming problem, that is
\begin{align}
&\phi(\omega,\overline{\omega}) = -\omega^{\mathrm{T}} \mathbf{Q(\overline{\omega})} \; \omega + \mathbf{b}^{\mathrm{T}} \omega \rightarrow \max_{\omega \in \Delta}.  \label{iteration}
\end{align}
This problem    can now be solved iteratively substituting each time the solution obtained in the previous iteration instead of $\overline{\omega}$.

\begin{exam} \label{exam1} {\rm
In this example we illustrate the calculation of the function $\mathbf{Q}(\omega) $ under  several distributional assumptions.
\begin{itemize}
\item[(1)]  \cite{ucibog2005}   considered the  regression model   with normal distributed heteroscedastic errors, that is
$$
f(y,x,\theta) = \frac {1}{\sqrt{2 \pi v^2 (x,\theta)}} \exp \Bigl ( - \frac {(y- \eta(x,\theta))^2}{2 v^2(x,\theta)} \Bigr),
$$
where $\eta(x,\theta)$ and $v^2(x, \theta) $ denotes the expectation and variance of the response at experimental condition $x$.
In this case the  Kullback-Leibler distance between the two densities $f_i$ and $f_j$ is given by
\begin{align*}
I_{i,j} ({x},\overline{\theta}_i ,{\theta}_{i,j}) =  \frac{\left[ \eta_i(x,\overline{\theta}_i) - \eta_j(x,\theta_{i,j}) \right]^2}{v_j^2(x,\theta_{i,j})} + \frac{v_i^2(x,\overline{\theta}_{i})}{v_j^2(x,\theta_{i,j})} + \log\left\{\frac{v_j^2(x,\theta_{i,j})}{v_i^2(x,\theta_i)}\right\}-1,
\end{align*}
and a straightforward calculation gives for the function $\bar g$ in \eqref{gbar} the representation 
\begin{align*}
\overline{g}(\omega)
&= \sum_{i,j=1}^{\nu} \min_{\alpha_{i,j}} \left[ \alpha_{i,j}^T \mathbf{J}_{i,j}^T \mathbf{\Omega}_i \mathbf{J}_{i,j} \alpha_{i,j} + 2 \omega^T \mathbf{R}_{i,j} \alpha_{i,j} + \mathbf{b}_{i,j}^T \omega \right],
\end{align*}
where $\mathbf{\Omega}_i = \mbox{diag} \left( \omega_1, \dots, \omega_n \right),$ 
\begin{align*}
&s_{i,j}(x,\theta_{i,j}) = \frac{v_i^2(x,\theta_{i})}{v_j^2(x,\theta_{i,j})} + \log\left\{\frac{v_j^2(x,\theta_{i,j})}{v_i^2(x,\theta_i)}\right\} ; \;  h_{i,j}(x,\theta_{i,j}) = \frac{\eta_i(x,\theta_i) - \eta_j(x,\theta_{i,j})}{v_j(x,{\theta}_{i,j})}; \\
&\mathbf{J}_{i,j} = \left( \frac{\partial h_{i,j}(x_k,\theta_{i,j})}{\partial \theta_{i,j}}\Big|_{\theta_{i,j} = \widehat{\theta}_{i,j}} \right)_{k = 1,\dots,n};  \\
&\mathbf{R}_{i,j} = \left( h_{i,j}(x_k,\widehat{\theta}_{i,j})\frac{ \partial h_{i,j}(x_k,\theta_{i,j})}{\partial \theta_{i,j}}\Big|_{\theta_{i,j}=\widehat{\theta}_{i,j}} + \frac{1}{2}  \frac{\partial s_{i,j}(x_k,\theta_{i,j})}{\partial \theta_{i,j}}\Big|_{\theta_{i,j} = \widehat{\theta}_{i,j}}\right)_{k = 1,\dots,n}; \\
& \mathbf{b}_{i,j} = \left( I_{i,j}(x_k,\overline{\theta}_i,\widehat{\theta}_{i,j}) \right)_{k = 1,\dots,n}
\end{align*}

\item[(3)]    \cite{loptomtra2007} considered the regression model \eqref{1.1} with log-normal distribution
with parameters  $ \mu (x,\theta)$ and $\sigma^2(x,\theta)$. This means that the  mean and the variance
are given by
 \begin{eqnarray*}
 \mathbb{E}[Y] &=& \eta(x,\theta) = \exp \Bigl \{ \frac {\sigma^2(x,\theta)}{2} + \mu (x,\theta) \Bigr \}, \\  
  \mbox{Var}(Y) &=& v^2(x,\theta) =  \eta^2 (x,\theta) \{ \exp \{ \sigma^2(x,\theta) \} - 1  \}~,
 \end{eqnarray*}
 respectively,
and the density  of the response $Y$ is given by 
 \begin{align*}
 f(y,x, \theta) = \frac{1}{x \sqrt{2 \pi} \sigma(x,\theta)} \exp\Big \{-\frac{\left\{ \log(y) - \mu(x,\theta) \right\}^2}{2 \sigma^2(x,\theta)}\Big\}
 \end{align*}
 In the paper~\cite{loptomtra2007} it was shown that the Kullback-Leibler distance between
 two  log-normal densities with parameters $\mu_\ell  (x,\theta_\ell) $ and $\sigma_\ell^2(x,\theta_\ell)$
 ($\ell=i,j$) is given by
\begin{align} \label{fin flop}
I_{ij} (x,{\theta}_i, {\theta}_{i,j})  =  \frac{1}{2}\Bigl\{s_{i,j}(x,\theta_{i,j}) + \frac{\left[\mu_i(x,\theta_i)-\mu_j(x,\theta_{i,j})\right]^2}{\sigma_i^2(x,\theta_i)}-1\Bigr\},
\end{align}
where
\begin{align*}
 s_{i,j}(x,\theta_{i,j}) = \log\Bigl[\frac{\sigma_i^2(x,\theta_i)}{\sigma_j^2(x,\theta_{i,j})}\Bigr] + \frac{\sigma_j^2(x,\theta_{i,j})}{\sigma_i^2(x,\theta_i)}, 
\end{align*}
and 
\begin{align*}
\sigma_i^2(x,\theta_i) &= \log\left[1+v^2_i(x,\theta_i)/\eta_i^2(x,\theta_i)\right],\\
\mu_i(x,\theta_i) &= \log\left[\eta_i(x,\theta_i)\right] - \sigma^2_i(x,\theta_i)/2.
\end{align*}
Now a straightforward calculation gives for the function $\bar g$ in \eqref{gbar} 
 the representation
\begin{align*}
\overline{g}(\omega) &= \frac{1}{2} \sum_{i,j=1}^{\nu} \min_{\alpha_{i,j}} \left[ \alpha_{i,j}^T \mathbf{J}_{i,j}^T \mathbf{\Omega}_i \mathbf{J}_{i,j} \alpha_{i,j} - 2 \omega^T \mathbf{R}_{i,j} \alpha_{i,j} + \mathbf{b}_{i,j}^T \omega \right],
\end{align*}
where $\mathbf{\Omega}_i = \mbox{diag} \left( \omega_1, \dots, \omega_n \right)$,
\begin{align*}
\mathbf{J}_{i,j} &= \left( \frac{\frac{\partial \mu_j(x_k,\theta_{i,j})}{\partial \theta_{i,j}}\Big|_{\theta_{i,j} = \widehat{\theta}_{i,j}}}{\sigma_i(x_1,\overline{\theta}_i)} \right)_{k = 1,\dots,n}; \\
\mathbf{R}_{i,j} &= \left( \frac{\big[\mu_i(x_k,\overline{\theta}_i)-\mu_j(x_k,\widehat{\theta}_{i,j})\big] \frac{\partial \mu_i(x_k,\theta_{i,j})}{\partial \theta_{i,j}}\big|_{\theta_{i,j}=\widehat{\theta}_{i,j}}}{\sigma^2_i(x_k,\overline{\theta}_i)} - \frac{1}{2} \frac{\partial s_{i,j}(x_k,\theta_{i,j})}{\partial \theta_{i,j}}\Big|_{\theta_{i,j} = \widehat{\theta}_{i,j}}  \right)_{k = 1,\dots,n}; \\
\mathbf{b}_{i,j} & = \left( I_{i,j}(x_k,\overline{\theta}_i,\widehat{\theta}_{i,j}) \right)_{k = 1,\dots,n}
\end{align*}
\end{itemize}
}
 \end{exam}

\subsection{A gradient method} \label{sec32}

In this section we describe a specialized gradient method for second step of  Algorithm~\ref{algorithm:new}
function 
To be precise we introduce theunctions
\begin{align*}
v_k(\omega) = \sum_{i,j = 1}^{\nu} p_{i,j} I_{i,j} ( x_k,{\overline{\theta}_{i}},\widehat{\theta}_{i,j}(\omega)) , \;  k = 1,\dots,n,
\end{align*}
where 
$\widehat{\theta}_{i,j}  = \widehat{\theta}_{i,j} (\omega) $ is defined in \eqref{thetatilde}.
Next we  iteratively calculate a sequence of vectors  $(\omega_{(\gamma)})_{\gamma =0,1,\ldots}$
starting with a vector 
$\omega_{(0)} = \overline{\omega}$ (for example equal weights).  For  
 ${\omega}_{(\gamma )} =({\omega}_{(\gamma ),1}, \ldots,{\omega}_{(\gamma ),n})$
we  determine indices
 $\overline{k} $ and  $ \underline{k}$ corresponding to
$ \max_{1 \leq k \leq n} v_k(\omega_{(\gamma)})$ and
$\min_{1 \leq k \leq n} v_k(\omega_{(\gamma)})$, respectively, and define
\begin{align} \label{argmax}
\alpha^* = \arg \max_{0 \leq \alpha \leq \omega_{(\gamma),\underline{k}}} g(\overline{\omega}_{(\gamma)}(\alpha)),
\end{align}
where the vector  $\overline{\omega}_{(\gamma )} (\alpha)
=(\overline{\omega}_{(\gamma ),1}(\alpha), \ldots,\overline{\omega}_{(\gamma ),n}(\alpha))$
is  given by
$$
\overline{\omega}_{(\gamma),i} (\alpha)= \left\{ \begin{array}{ll}
\omega_{(\gamma),i}  + \alpha & \mbox{ if } i=\overline{k}\\
\omega_{(\gamma),i}  -  \alpha & \mbox{ if }  i= \underline{k}\\
\omega_{(\gamma),i}   & \mbox{ else } \\
\end{array}
\right.
$$
The vector ${\omega}_{(\gamma+1)} $ of the next iteration is then defined by  $
{\omega}_{(\gamma+1)}  = \overline{\omega}_{(\gamma)} (\alpha^*).$
 It  follows by similar arguments as in  \cite{detmelguc2015} 
 that the generated sequence of vectors converges to a maximizer of the function $g$.

\section{Implementation and  numerical examples }\label{sec4}

In this section we illustrate the new algorithms calculating Bayesian
KL-optimal discriminating designs for several models with non-normal errors. 
We begin giving a few more details regarding the implementation.

\begin{itemize}
\item[(1)]
As pointed out in Section \ref{sec2}
 a  Bayesian KL-optimality  criterion is   reduced to a local criterion of the form \eqref{2.4} for a large
 number of model comparisons. For illustration purposes, consider the
 criterion  \eqref{2.6}, where  $\nu=2$, $p_{1,2}=1, p_{2,1}=0$ and the prior for  the parameter $\theta_1$ puts masses
 $\tau_{1}, \ldots \tau_{\ell}$ at the points $\lambda_{1},\ldots ,\lambda_{\ell}$. This criterion can  be rewritten as
 a local  criterion of the form \eqref{2.4}, i.e.
\begin{align}
\mathrm{KL}_{\mathrm{P}}(\xi) = \sum_{i,j=1}^{\ell+1}  \widetilde{p}_{i,j} \inf_{\theta_{i,j} \in \Theta_j} \int_{\mathcal{X}}
I_{i,j} (x,\theta_i,\theta_{i,j}) \xi(dx),
\end{align}
where $\widetilde p_{1,\ell +1} =\tau_1, \ldots , \widetilde p_{1,\ell +1} =\tau_\ell$ and all other weights $\widetilde p_{i,j}$ are $0$.
The extension of this approach to
more than two models is easy and left to the reader. 
\item[(2)]
In Step 1 of Algorithm~\ref{algorithm:new}   all local maxima of the function $\Psi(x,{\xi_s})$ 
are aded as possible support points of the design in the next  iteration.
In order to  avoid the problem of accumulating  too many
support points we remove in each iteration those
points with a  weight  smaller than $m^{0.25}$, where $m= 2.2\times 10^{-16} $ is 
the working precision  R. 
\item[(3)] 
In the implementation of the quadratic programming method for Step 2 of  Algorithm~\ref{algorithm:new} 
(see Section \ref{sec31})
we  perform only a few iterations such  that an improvement  compared to  the starting design
is obtained. This speeds up the convergence of the procedure substantially
without affecting the convergence in all examples under consideration.
\item[(4)] In the   implementation of the gradient method for Step 2 of  Algorithm~\ref{algorithm:new} 
(see Section \ref{sec32})
we use a  linearization of the optimization problem in order to improve the speed of the procedure. 
\end{itemize}
We are now ready to demonstrate the advantages of 
 the new method in several  examples  calculating Bayesian KL-optimal discriminating designs.
For the sake of brevity  we restrict ourselves to the case of non-linear regression models, where the response  has
a  log-normal distribution with parameters $\mu (x,\theta) $ and $\sigma^2(x,\theta)$ as described in Example \ref{exam1}.

\begin{exam}\label{exam2}{\rm
Our first example refers the problem of determining  local  KL-optimal designs for a situation investigated
by \cite{loptomtra2007}. Motivated by  pharmacokinetic  practice [see \cite{lindsey2001,crawley2002}]
these  authors determined local KL-optimal designs for
 two log-normal models with mean functions
\begin{align} \label{mm}
\eta_1(x,\theta_1) = \frac{\theta_{1,1} x}{\theta_{1,2} + x} + \theta_{1,3} x ,\qquad \eta_2(x,\theta_2) = \frac{\theta_{2,1} x}{\theta_{2,2} + x}
\end{align}
on the interval $\mathcal{X} = [0.1,5]$. They assumed equal and constant variances and considered model $\eta_1$
with parameter  $\theta_1 = (1,1,1)$ as fixed. This corresponds to the choice $\nu=2$ and $p_{1,2} = 1, \; p_{2,1} = 0$
in the criterion \eqref{2.4}. In Table \ref{tab1} we resent the optimal design calculating by the new algorithms for various choices
of the mean and variance function, that is
\begin{equation} \label{modtab1}
\begin{array}{ll}
(1) &   v^2_1(x,\theta_1)=v^2_2(x,\theta_2)=1  \\
(2) &\sigma^2_1(x,\theta_1)=\sigma^2_2(x,\theta_2)=1 \\
(3) & v^2_i(x,\theta_i) = \exp(\eta_i(x,\theta_i))
\end{array}
\end{equation}
All designs have an efficiency that is at least $0.999$, and we have used three methods
for the calculation of the local KL-optimal design. The first procedure is a classical exchange type method as
proposed by \cite{loptomtra2007}. The other methods are the two versions of the new Algorithm \ref{algorithm:new}
with the modifications described  in Section \ref{sec31} (quadratic programming) and \ref{sec32} (gradient method).
For the case (2) of equal variances in  \eqref{modtab1} the corresponding function  in \eqref{fin flop}
simplifies to  $ (\mu_1(x,\bar \theta_1)-\mu_2(x,\theta_{1,2}))^2 $. Consequently, one can use the procedure for the special case of a normal
distributed response  developed in \cite{detmelguc2015}, where   $\mu_i(x,\theta_i) = \log \eta_i(x,\theta_i)$  ($i=1,2$),
which works significantly faster. It should be noted that the calculated designs slightly differ  from those in
\cite{loptomtra2007}. \\
 In Table \ref{tab1} we also show the computation time (CPU  time in seconds  on a standard PC with an intel core i7-4790K processor)
 for the different methods.
We observe that the methods developed in this paper work
substantially faster than the exchange type algorithm proposed in  \cite{loptomtra2007}.  For example, the new
gradient methods are between $5$ and $30$ times faster, while the quadratic programming approach yield to a procedure
which is between $25$ and $120$ times faster than the classical exchange type algorithm. In the case of two competing models
the exchange type algorithm is still finding the local KL-optimal discriminating design in a reasonable time, but the difference become
more important if a discriminating design has to be found for more than two competing models or if a Bayesian  KL-optimal design has to be determined.
Some of these situations are discussed in the following examples.
 \begin{table}[h]
\begin{center}
\begin{tabular}{|c|c|c|c|c|}
\hline
case & KL-opt. design & AF  & grad & quad \\
\hline
(1) & $\begin{matrix} 0.130 & 2.501 & 5.000 \\ 0.489 & 0.378 & 0.133\end{matrix}$ & 7.15 & 0.56 & 0.06 \\
\hline
(2) & $\begin{matrix} 0.100 & 1.569 & 5.000 \\ 0.294 & 0.500 & 0.206\end{matrix}$ & 2.74 & 0.52 & 0.01 \\
\hline
(3) & $\begin{matrix} 0.100 & 1.218 & 5.000 \\ 0.326 & 0.510 & 0.164\end{matrix}$ & 10.87 & 0.33 & 0.08 \\
\hline
\end{tabular}
\end{center}
 \caption{\it \label{tab1} Local  KL-optimal discriminating designs for the models in   \eqref{mm}. The responses are  log-normal
 distributed  with
 different  specifications of  the mean and variance - see \eqref{modtab1}.  Column 3 - 5 show the computation time of the different algorithms
 (exchange type Algorithm \ref{algorithm:AtkinsonFedorov}   (AF) proposed in
   \cite{loptomtra2007} and Algorithm \ref{algorithm:new} with a gradient (grad) and quadratic programming method  (quad) in Step 2).}
\end{table}
}
\end{exam}

\begin{exam}\label{exam3}{\rm In our second example we calculate Bayesian KL-optimal discriminating design for
two competing  exponential models
\begin{align}
&\eta_1(x,\theta_1) = \theta_{1,1} - \theta_{1,2} \exp(-\theta_{1,3} x^{\theta_{1,4}});  \label{example1}\\
&\eta_2(x,\theta_2) = \theta_{2,1} - \theta_{2,2} \exp(-\theta_{2,3} x).\nonumber
\end{align}
on the interval $[0,10]$, where model $\eta_1$ is again fixed. Discriminating  designs for these models have been
determined by  \cite{detmelguc2015} under the assumption of a normal distribution, and we will now investigate how the designs
change for the log-normal distributed responses with mean  and variance   specified by \eqref{modtab1}. Following these authors we considered independent
  prior distributions supported at the points
  \begin{equation} \label{sup1}
  \mu_j + \frac {\sqrt{0.3}(i-3)}{2}, \qquad \qquad i=1,\dots,5 \, ; \quad j=3,4 \, ,
  \end{equation}
for the parameters $\bar \theta_{1,3}$ and $\bar \theta_{1,4}$  where $\mu_3 = 0.8, \ \mu_4 = 1.5$.
The corresponding weights at these points are
  proportional  (in both cases) to
  \begin{equation} \label{wei1}
  \frac {1}{\sqrt{2 \pi \cdot 0.3}} \exp \Bigl( - \frac {(i-3)^2}{8} \Bigr); \qquad \qquad i=1,\dots, 5\, .
  \end{equation}
Note that  the optimal discriminating designs do not depend on the linear parameters of  $\eta_1$ , for which we have chosen
  as    $\bar \theta_{2,1}=2$ and $\bar \theta_{2,2}=1 $. 
  
  \begin{figure}[t]
\centering
 {\includegraphics[width=52mm]{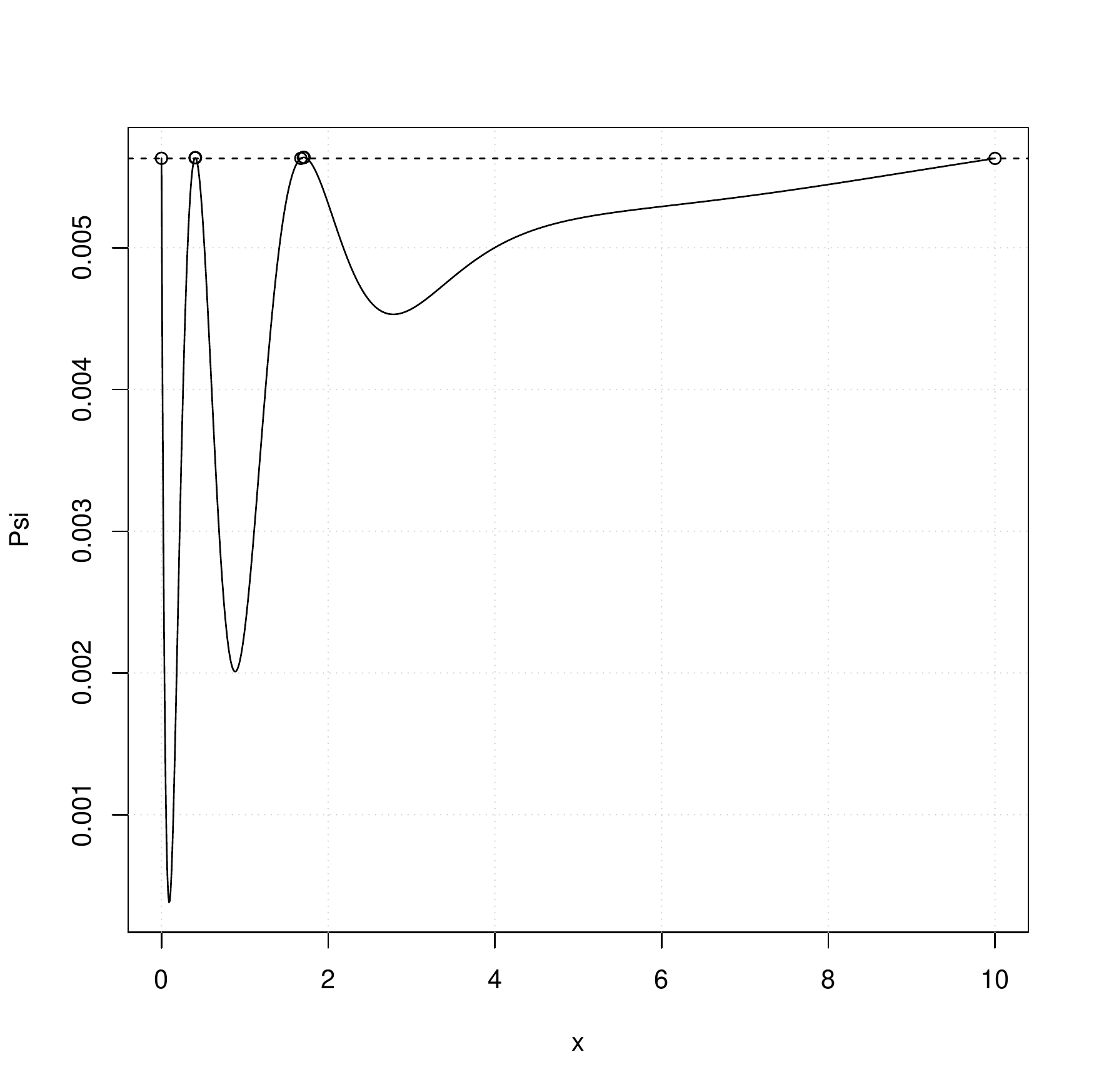}} ~~~
 {\includegraphics[width=52mm]{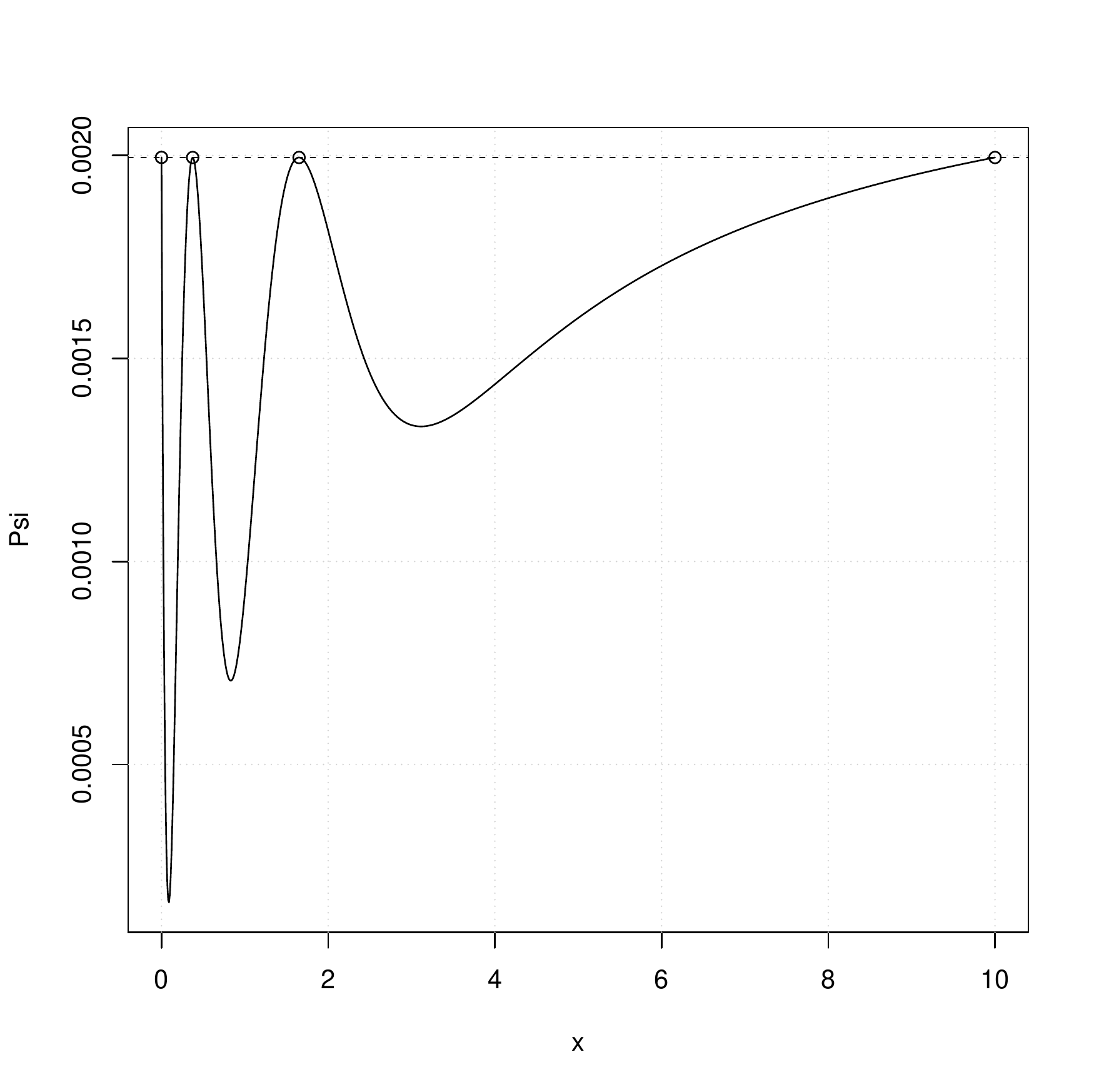}}~~~
 {\includegraphics[width=52mm]{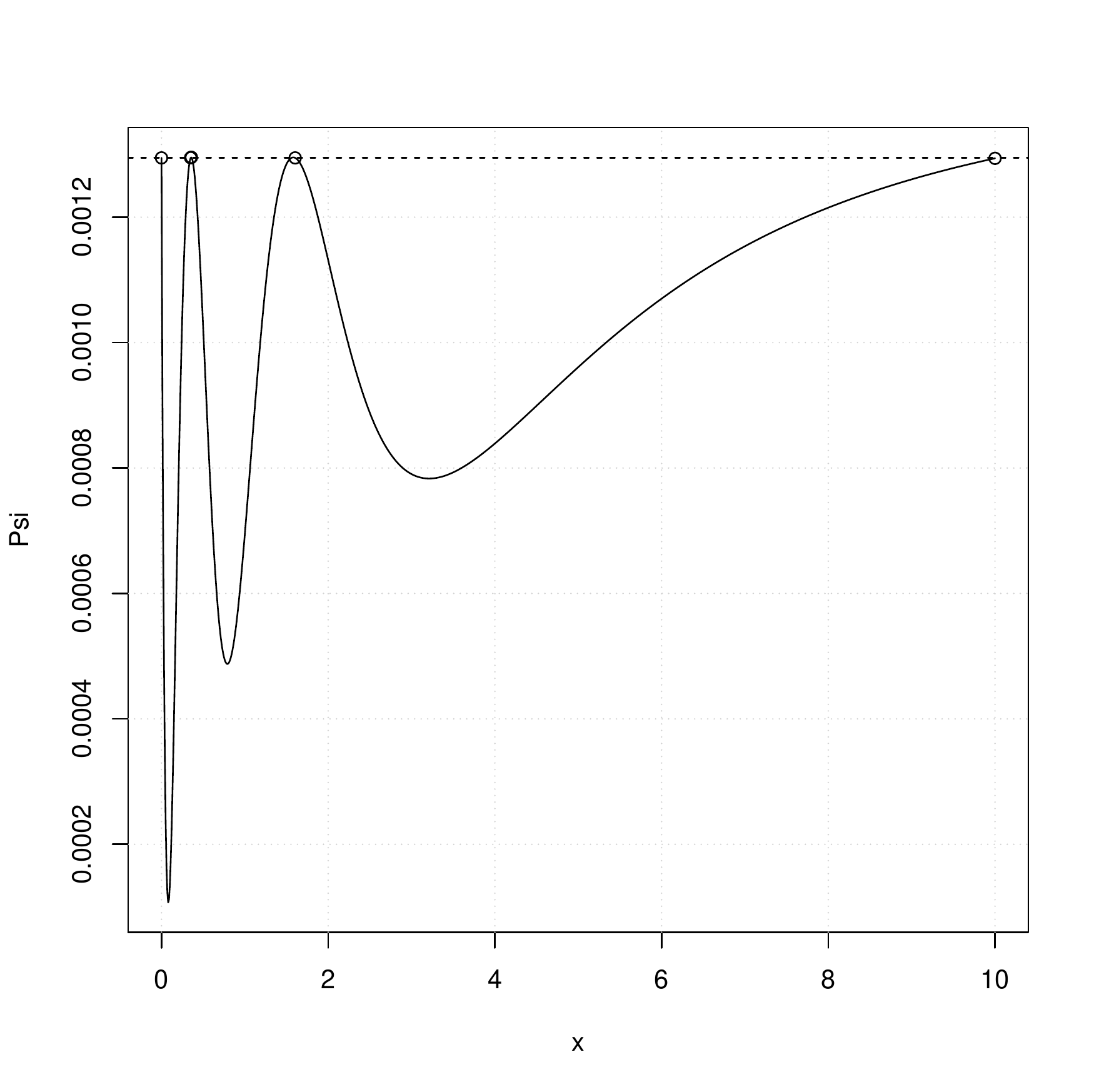}} \\
~~~(1) ~~~~~~~~~~~~~~~~~~~~~~~~~~~~~~~~~~~~~~~~~(2) ~~~~~~~~~~~~~~~~~~~~~~~~~~~~~~~~~~~~~~~~~(3) ~
 \caption{\it \label{fig1} The function on the left hand side of inequality
\eqref{equiv} in the equivalence Theorem \ref{thm1} for the numerically calculated Bayesian KL-optimal discriminating designs.
The  competing regression models are given in \eqref{example1}.}
\end{figure}
   The Bayesian KL-optimal discriminating designs for log-normal distributed responses are displayed in Table \ref{tab2} for the
   different specifications of the mean and variance in \eqref{modtab1}.  In Figure \ref{fig1} we show the
    function on the left hand side of inequality \eqref{equiv} in the equivalence Theorem \ref{thm1}.
   Comparing the computational times in Table \ref{tab2} we observe again that using
   quadratic programming  in Step 2  of  Algorithm \ref{algorithm:new}
    is substantially faster than the gradient method.  \\
    It might be of interest to compare the Bayesian optimal discriminating designs for the various
    log-normal distributed responses    with the design for normal distributed responses determined in \cite{detmelguc2015}.
    This design is supported at the five(!) points $0.000$, $ 0.452$, $  1.747$, $  4.951$ and $  10.000$
    with masses $ 0.207$, $  0.396$, $  0.292$, $  0.003$  and  $  0.102$, respectively. The efficiencies
     $$\mathrm{Eff}_{KL_P}^{(j)}(\xi_{(i)}^*) = \frac{KL_P^{(j)}(\xi_{(i)}^*)}{\sup_{\eta} KL_P^{(j)}(\eta)}.$$
     under misspecification of the distribution of the response are depicted in Table \ref{tab3}. For example,
     the efficiency of the design $\xi^*_{(0)}$ calculated under  the assumption homoscedastic
     normal distributed responses in the model with log-normal distributed responses in \eqref{modtab1}(2)
     is given by $95.3\%$. We observe that the Bayesian optimal discriminating designs calculated for normal
     distributed responses are rather robust and have good efficiencies  for the log-normal distribution.
   \begin{table}[h]
\begin{center}
\begin{tabular}{|c|c|c|c|c|c|}
\hline
\eqref{modtab1} & design & AF & grad & quad \\
\hline
(1) & $\begin{matrix} 0 & 0.406 & 1.706 & 10 \\  0.186 & 0.418 & 0.289 & 0.107 \end{matrix}$ & 298.37 & 44.36 & 3.7 \\
\hline
(2) & $\begin{matrix} 0 & 0.374 & 1.650 & 10 \\  0.189 & 0.397 & 0.311 & 0.103 \end{matrix}$ & 390.44 & 7.39  & 2.39 \\
\hline
(3) & $\begin{matrix} 0 & 0.356 & 1.604 & 10 \\  0.186 & 0.394 & 0.313 & 0.107 \end{matrix}$ & 570.45 & 39.19 & 4.42 \\
\hline
\end{tabular}
\end{center}
 \caption{\it \label{tab2}  Bayesian  KL-optimal discriminating designs for the models in  \eqref{example1}. The responses are  log-normal
 distributed  with
 different  specifications of  the mean and variance - see \eqref{modtab1}. The prior distribution is
 defined by \eqref{sup1} and \eqref{wei1}.
 Column three and four   show the computation time of the new  Algorithm \ref{algorithm:new}
 proposed in this paper with a gradient (grad) and quadratic programming method  (quad) in Step 2.}
\end{table}

\begin{table}[h]
\begin{center}
\begin{tabular}{|c|c|c|c|c|}
\hline
 & (0)  & (1)  &(2)  & (3)  \\
\hline
(0)     & 1 & 0.978 & 0.953 & 0.908 \\
\hline
(1) & 0.981 & 1 & 0.988 & 0.966 \\
\hline
(2)  & 0.951 & 0.987 & 1 & 0.992 \\
\hline
(3) & 0.923 & 0.970 & 0.996 & 1 \\
\hline
\end{tabular}
\end{center}
 \caption{\it \label{tab3}  Efficiencies of  Bayesian  KL-optimal discriminating designs for the models in  \eqref{example1}
 under different distributional assumptions for the responses.  (0): homoscedastic normal distribution; (1) - (3):
   log-normal  distribution   with different  specifications of  the mean and variance - see \eqref{modtab1}.}
\end{table}
}
\end{exam}

\begin{table}[t]
\begin{center}
\begin{tabular}{|c|c|c|c|c|}
\hline
\eqref{modtab2} & design & AF & grad & quad \\
\hline
(1) & $\begin{matrix} 0.759 & 67.32 & 248.6 & 500 \\  0.419 & 0.156 &  0.233 & 0.192\end{matrix}$ & 1674.14 & 679.52 & 48.91 \\
\hline
(2) & $\begin{matrix} 0 & 58.9 & 220.6 & 500 \\  0.200 & 0.354 &  0.247 & 0.199\end{matrix}$ & - &  255.03 & 33.42 \\
\hline
(3) & $\begin{matrix} 0 & 33.12 & 78.0 & 161.6 & 215.7 & 500 \\ 0.279 & 0.092 & 0.225 & 0.003 & 0.224 & 0.177 \end{matrix}$ & 2382.64 & 631.53 & 82.33 \\ 
\hline 
\end{tabular}
\end{center}
 \caption{\it \label{tab4}  Bayesian  KL-optimal discriminating designs for the 
 competing dose response models in  \eqref{example2}. The responses are  log-normal
 distributed  with
 different  specifications of  the mean and variance - see \eqref{modtab2}. The prior distribution is
 a uniform distribution on $81$ points as specified 
in   \eqref{uni1}.
 Column three, four and five   show the computation time of the exchange type algorithm  (AF), the new  Algorithm \ref{algorithm:new}
 proposed in this paper with a gradient (grad) and quadratic programming method  (quad) in Step 2.}
\end{table}
\begin{exam}\label{exam4}{\rm  Our final example refers to the construction of Bayesian KL-optimal discriminating designs
for several dose response curves, which have been recently proposed by \cite{pinbrebra2006}
for modeling  the dose response relationship of a Phase II clinical trial, that is
\begin{align}
&\eta_1(x, \theta_1) = \theta_{1,1} + \theta_{1,2} x ;  \nonumber \\
&\eta_2(x, \theta_2) = \theta_{2,1} + \theta_{2,2} x (\theta_{2,3} - x) ; \label{example2}  \\
&\eta_3(x, \theta_3) = \theta_{3,1} + \theta_{3,2} x / (\theta_{3,3} + x) ; \nonumber\\
&\eta_4(x, \theta_4) = \theta_{4,1} + \theta_{4,2} / (1 + \exp(\theta_{4,3} - x) / \theta_{4,4}) ; \nonumber
\end{align}
where the designs space (dose range) is given by the interval
$\mathcal{X}=[0,500]$. In this reference  some prior information  regarding the parameters for the
models is also provided., that is
\begin{align*}
\overline{\theta}_1= (60, 0.56), \; \overline{\theta}_2= (60, 7/2250, 600), \; \overline{\theta}_3 = (60, 294, 25), \; \ \overline{\theta}_4=
(49.62, 290.51, 150, 45.51).
\end{align*}
 \cite{detmelguc2015} determined Bayesian KL-optimal discriminating designs for these 
models under the assumption of normal distributed responses,
where they used   $p_{i,j}=1/6$, $(1\leq j<i \leq 4 )$ and they assumed that there exist only uncertainty 
for the parameter $\theta_4$. We will now consider similar problems for log-normal distributed responses,
where the  prior distribution is a uniform distribution at $81$ points in $\R^4$, that is
\begin{align} \label{uni1}
(49.62+c_1, 290.51+c_2, 150+c_3, 45.51+c_4)
\end{align}
with   $c_1,c_2,c_3,c_4 \in \{  -20,0,45\} $. Note that we cannot use the prior distribution considered in 
 \cite{detmelguc2015} because this would yield a negative mean $\eta_i(x,\theta_i)$.
  The resulting Bayesian optimality criterion \eqref{2.6}  consist of $246$ model comparisons 
  and Bayesain KL--optimal discriminating designs are depicted in Table \ref{tab4} for the cases
\begin{equation} \label{modtab2}
\begin{array}{ll}
(1) &   v^2_1(x,\theta_1)=1~,~ i = 1,2,3,4 ;\\
(2) &\sigma^2_i(x,\theta_i)=1 ~,~ i = 1,2,3,4 ;\\ 
(3) &v_i^2(x,\theta_i)=\exp( \eta_i(x,\theta_i)/100)~,~ i = 1,2,3,4 .
\end{array}
\end{equation}
All calculated designs have at least efficiency $99.9\%$   and the  corresponding plots of the equivalence Theorem
\ref{thm1} are shown in Figure \ref{fig2}. In the models specified by \eqref{example2} all new algorithms were able to find the
Bayesian KL-optimal discriminating design, where the exchange type algorithm failed in the case \eqref{modtab2}(2). 
Moreover, in the other cases  the new methods are substantially faster than the exchange type method. For example, the gradient method 
yields only $25\%-30\%$ of the computational time, while the quadratic programming approach is about $30-35$ times faster.

  \begin{figure}[h]
\centering
 {\includegraphics[width=52mm]{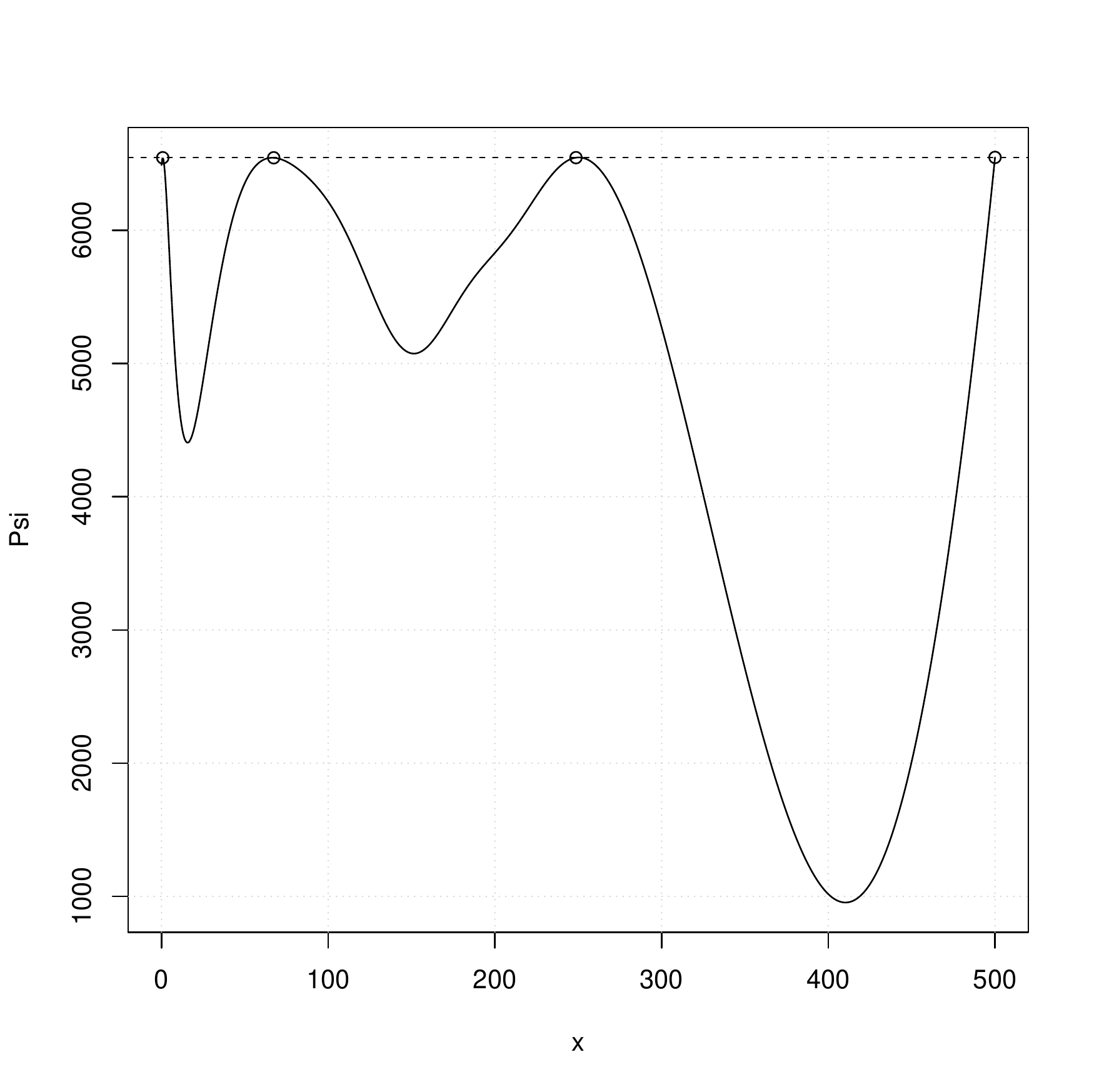}} ~~~
 {\includegraphics[width=52mm]{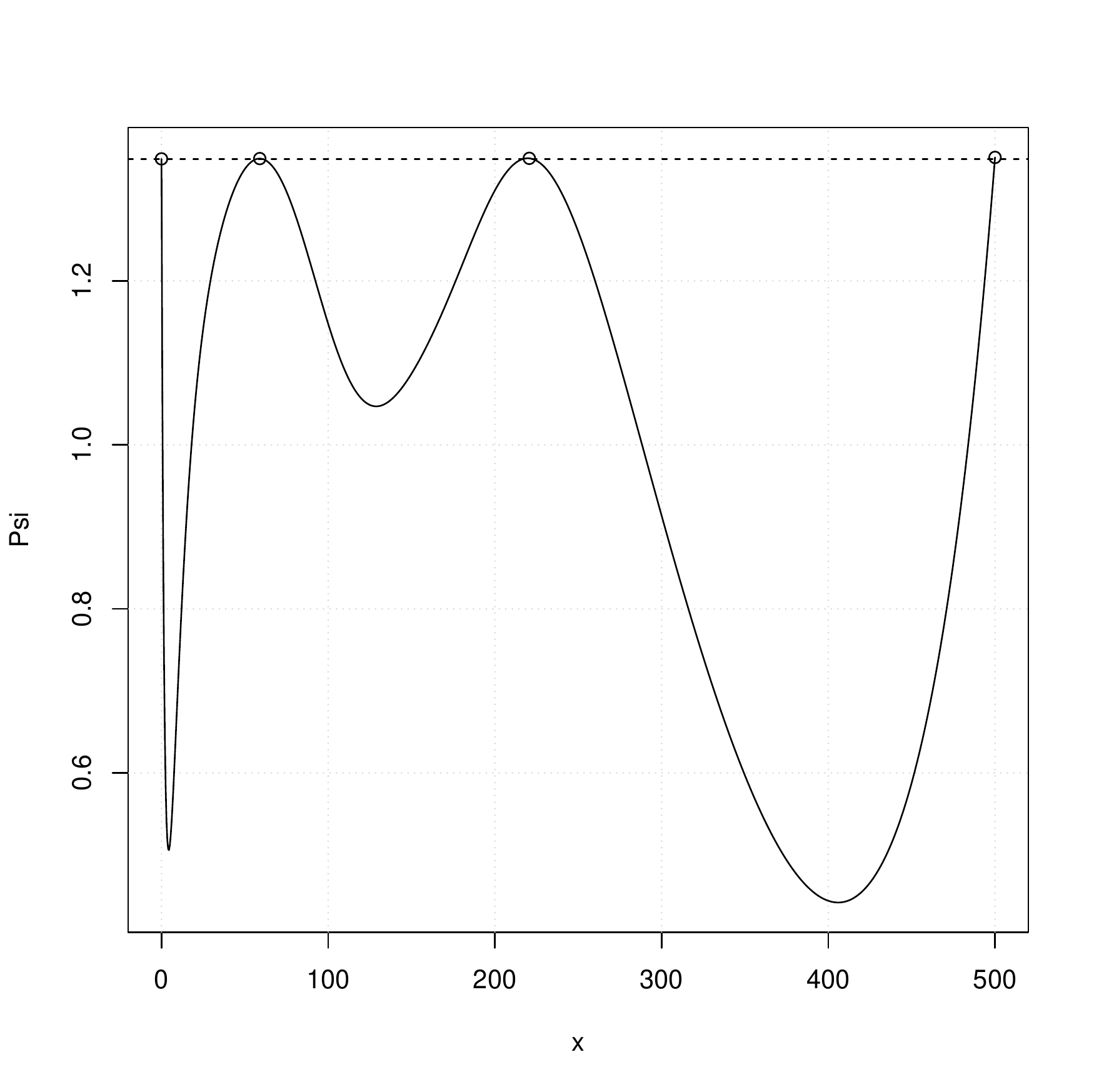}}~~~
 {\includegraphics[width=52mm]{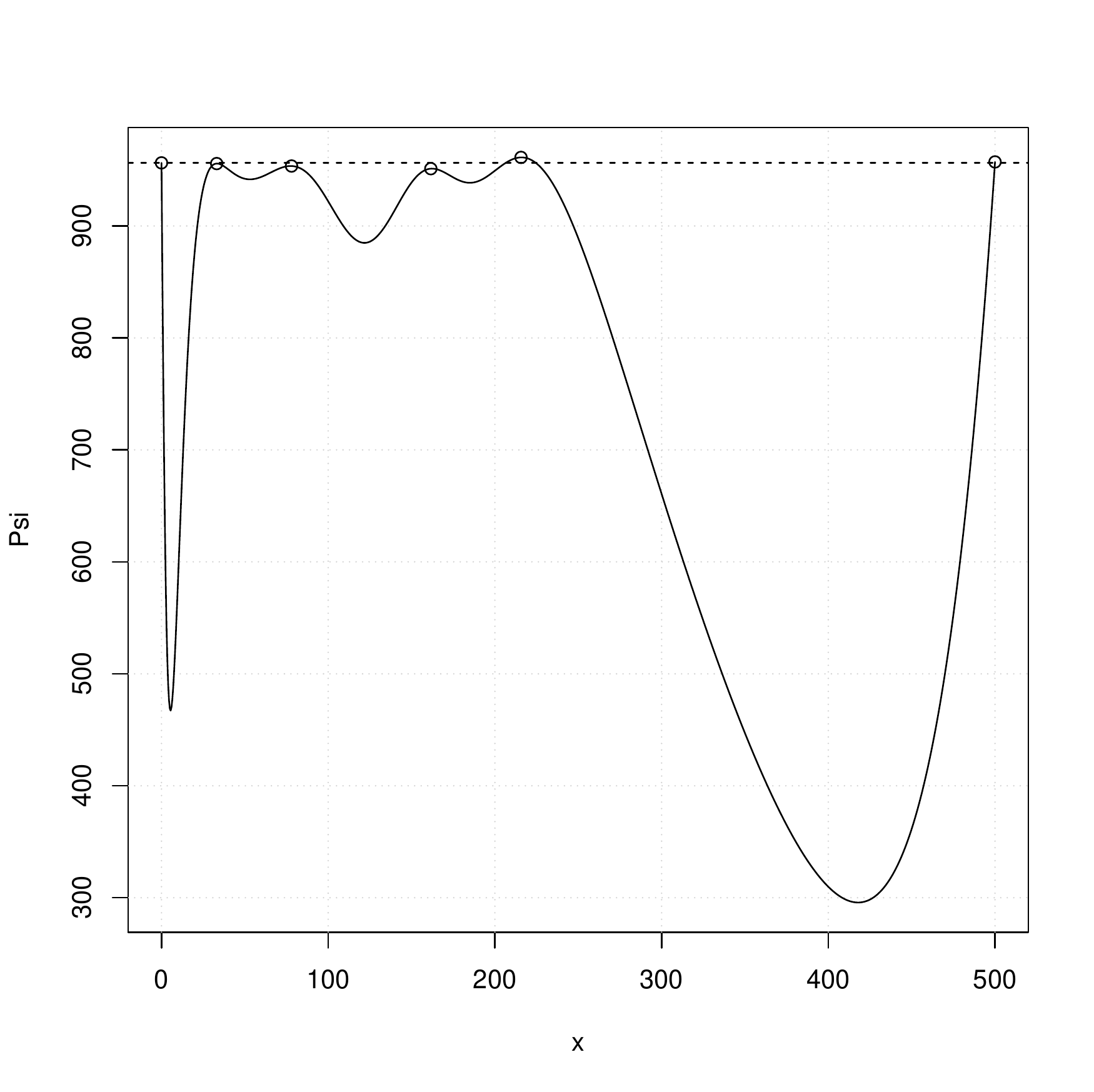}} \\
~~~(1) ~~~~~~~~~~~~~~~~~~~~~~~~~~~~~~~~~~~~~~~~~(2) ~~~~~~~~~~~~~~~~~~~~~~~~~~~~~~~~~~~~~~~~~(3) ~
 \caption{\it \label{fig2} The function on the left hand side of inequality
\eqref{equiv} in the equivalence Theorem \ref{thm1} for the numerically calculated Bayesian KL-optimal discriminating designs.
The  competing regression models are given in \eqref{example2} and the scenarios for log-normal distribution specified in \eqref{modtab2}. }
\end{figure}
}
\end{exam}

\section{Appendix: Proof of Lemma \ref{lem1}} \label{sec5}

By the construction of the function  $\overline{g}$ we have
$
\overline{g}(\omega) \leq g(\omega)
$  for all $ \omega \in \Delta$.
Let $\omega^* \in  \argmax_{\omega \in \Delta} g(\omega)$ and  define the function
\begin{align*}
\Psi_{i,j}(\theta_{i,j},\omega) = \sum_{k=1}^n \omega_k \int \log \left\{\frac{f_i(x_k,y,\overline{\theta}_i)}{f_j(x_k,y,{\theta}_{i,j})} \right\} f_i(x_k,y,\overline{\theta}_i) d \mu (dy)
\end{align*}
If $\widehat \theta_{i,j}  = \argmin_{\theta_{i,j}\in \Theta_j} \Psi_{i,j}(\theta_{i,j},\omega^*)$ is the miminizer of $ \Psi_{i,j}$ it follows from Assumption \ref{assum2}
that
\begin{align*}
\frac{\partial}{\partial \theta_{i,j}} \Psi_{i,j}(\theta_{i,j},\omega^*)\Big|_{\theta_{i,j} = \widehat{\theta}_{i,j}} = \mathbf{R}_{i,j}(\widehat \theta_{i,j}) \omega^* = 0,
\end{align*}
and we obtain
\begin{align*}
\widehat{\alpha}_{i,j} = \left( \int \mathbf{J}_{i,j}^T(y) \Omega \mathbf{J}_{i,j}(y) dy \right)^{-1} \mathbf{R}_{i,j}^T(\widehat{\theta}_{i,j}) \omega^* = 0
\end{align*}
Inserting this value in \eqref{gbar} gives $\overline{g}(\omega^*) = {g}(\omega^*)$, i.e.  $\max_{\omega \in \Delta}   \overline{g}(\omega) = \max_{\omega \in \Delta}   g(\omega)$. Now let $\omega^* = \argmax_{\omega \in \Delta}   \overline{g}(\omega)$. From the above equality it follows that $\widehat{\alpha}_{i,j} = 0$ and therefore $\mathbf{R}_{i,j}(\widehat \theta_{i,j})\omega^* = 0$, that is $\omega^* = \argmax g(\omega)$.
\hfill $\Box$

\bigskip

\bigskip

{\bf Acknowledgements.} Parts of this work were done during a visit of the
second author at the Department of Mathematics, Ruhr-Universit\"at
Bochum, Germany.
The authors would like to thank
M. Stein who typed this manuscript with considerable technical
expertise.
The work of H. Dette and V. Melas was supported by the Deutsche
Forschungsgemeinschaft (SFB 823: Statistik nichtlinearer dynamischer Prozesse, Teilprojekt C2).
The research of H. Dette reported in this publication was also partially supported by the National Institute of
General Medical Sciences of the National Institutes of Health under Award Number R01GM107639.
The content is solely the responsibility of the authors and does not necessarily
 represent the official views of the National
Institutes of Health. The work of V. Melas and R. Guchenko was also partially supported by  by St. Petersburg State University
(project "Actual problems of design and analysis for regression models", 6.38.435.2015).
\bigskip

\setstretch{1.25}
\setlength{\bibsep}{1pt}
\begin{small} \footnotesize
 \bibliographystyle{apalike}
\itemsep=0.5pt
\bibliography{model}
\end{small}
\end{document}